\documentstyle[aps,eqsecnum,epsf,preprint]{revtex}

\begin{document}
\preprint{\vbox{
\hbox{UCSD/PTH 98--17}
\hbox{hep-ph/9805416}
}}


\tightenlines

\title{
SU(3) symmetry breaking in hyperon semileptonic decays
}

\author{
Rub{\'e}n Flores-Mendieta,\footnote{On leave from Departamento de 
F{\'\i}sica, Centro de Investigaci{\'o}n y de Estudios Avanzados del IPN,
Apartado Postal 14-740, 07000, M{\'e}xico, Distrito Federal, Mexico}
Elizabeth Jenkins, and Aneesh V. Manohar}
\address{
Department of Physics, University of California at San Diego, La Jolla,
California 92093
}

\date{May 1998}

\maketitle

\begin{abstract}
\noindent 
Flavor $SU(3)$ symmetry breaking in the hyperon semileptonic decay
form factors is analyzed using the $1/N_c$ expansion. A detailed
comparison with experimental data shows that corrections to $f_1$ are
approximately 10\%, which agrees with theoretical expectations.
Corrections to $g_1$ are compatible with first-order symmetry breaking. A
fit to the experimental data allows one to predict the $g_1$ form factor
for $\Xi^0 \to \Sigma^+$ decay. The proton matrix element of the $T^8$
component of the axial current (which is equal to $3F-D$ in the $SU(3)$
symmetry limit) is found to be $\approx 0.34$--$0.46$.
\end{abstract}


\section{INTRODUCTION}

In order to determine the Cabibbo-Kobayashi-Maskawa (CKM) mixing matrix
from hyperon semileptonic decays (HSD), it is important to understand
flavor $SU(3)$ symmetry breaking effects in the hyperon $\beta$-decay form
factors.  At present, $V_{ud}$ can be precisely obtained from superallowed
$0^+ \rightarrow 0^+$ $\beta$-decays, but $V_{us}$ can be more reliably
determined from $K_{e3}$ decays than HSD~\cite{prd} because there are
larger uncertainties due to first-order symmetry breaking corrections in
the HSD axial-vector form factors than in kaon matrix elements.  Quark
model calculations including symmetry breaking corrections~\cite{don,sch}
predict that the vector form factor $f_1$ is smaller than its $SU(3)$
symmetric value. The value for $V_{us}$ obtained using this prediction is
incompatible with the one obtained from $K_{e3}$ decays~\cite{rfm},
$V_{us} = 0.2196 \pm 0.0023$.  However, a recent analysis of the data
favored $f_1$ larger than its $SU(3)$ symmetric value, which yields a
$V_{us}$ value consistent with the $K_{e3}$ extraction~\cite{rfm}. 

In this paper we incorporate $SU(3)$ symmetry breaking corrections into
the HSD form factors within the framework of the $1/N_c$ expansion of QCD.
The HSD form factors are analyzed in a combined expansion in $1/N_c$ and
$SU(3)$ flavor symmetry breaking.   We will base our analysis on the
formalism described in Ref.~\cite{dashen96}.  The organization of this
paper is as follows. Section~\ref{sec:HSD} gives a brief introduction to
the weak form factors relevant for HSD.  The $1/N_c$ expansion of the HSD
form factors is derived in Sec.~\ref{sec:OA}.  In order to make this
article self-contained, a brief description of the basics of the $1/N_c$
expansion is given as well. In Sec.~\ref{sec:FIT} we perform a detailed
comparison of the theoretical expressions with the available experimental
data~\cite{prd} for the decay rates, angular correlations and angular
spin-asymmetry coefficients of the octet baryons, and for the widths
(converted to axial-vector couplings through the Goldberger-Treiman
relation) of the decuplet baryons. Results and conclusions are presented
in Sec.~\ref{sec:RESULTS}. We find that the best fit values for $f_1$  are
larger than the $SU(3)$ symmetric values and yield a $V_{us}$ value
consistent with that obtained from $K_{e3}$ decays. The fit also gives a
good description of $SU(3)$ symmetry breaking for the axial form factor
$g_1$. The Fermilab KTeV collaboration will soon publish their initial
results for  $\Xi^0 \to \Sigma^+$ $\beta$-decay~\cite{swallow}. Our fit
predicts $f_1 \sim 1.1$ and $g_1 \sim 1.02$--$1.07$ for this decay.

\section{HYPERON SEMILEPTONIC DECAYS}\label{sec:HSD}
    
The low-energy weak interaction Hamiltonian for semileptonic decays is
given by
\begin{equation}
H_{W} = \frac{G}{\sqrt{2}} J_\mu L^\mu + {\rm h.c.} \, ,
\label{eq:hint}
\end{equation}
where the leptonic current
\begin{equation}
L^\mu = \overline \psi_e \gamma^\mu (1 - \gamma_5) \psi_{\nu_e} +
\overline \psi_\mu \gamma^\mu (1 - \gamma_5) \psi_{\nu_\mu} \, , 
\label{eq:lcurr}
\end{equation}
and the hadronic current expressed in terms of the vector ($V_\mu$)
and axial-vector ($A_\mu$) currents is 
\begin{eqnarray}
J_\mu & = & V_\mu - A_\mu \, , \nonumber \\
V_\mu & = & V_{ud} \overline u \gamma_\mu d + V_{us} \overline u
\gamma_\mu s \, , \label{eq:hcurr} \\
A_\mu & = & V_{ud} \overline u \gamma_\mu \gamma_5 d + V_{us} \overline u
\gamma_\mu \gamma_5 s \, . \nonumber
\end{eqnarray}
$G$ is the weak coupling constant, and $V_{ud}$ and $V_{us}$ are elements
of the CKM matrix. For definiteness, the notation and conventions of
Ref.~\cite{garcia85} are adopted in the present work.

The matrix elements of the hadronic current between spin-$1/2$ states can
be written as
\begin{eqnarray}
\langle B^\prime |V_\mu|B \rangle & = & V_{\rm CKM}\, 
\overline u_{B^\prime}(p_2) \left[ f_1(q^2)
\gamma_\mu + \frac{f_2(q^2)}{M_1} \sigma_{\mu\nu}q^\nu + 
\frac{f_3(q^2)}{M_1} q_\mu \right] u_B(p_1) \, , \label{eq:weak1} \\
\langle B^{\prime}|A_\mu|B \rangle & = & V_{\rm CKM}\, 
\overline u_{B^\prime}(p_2) \left[ g_1(q^2)
\gamma_\mu + \frac{g_2(q^2)}{M_1} \sigma_{\mu\nu}q^\nu + 
\frac{g_3(q^2)}{M_1} q_\mu \right] \gamma_5 u_B(p_1) \, , \label{eq:weak2}
\end{eqnarray}
where $u_B(p_1)$, $p_1$, $M_1$ [$\overline u_{B^\prime}(p_2)$, $p_2$,
$M_2$] are the Dirac spinor, the four-momentum, and the mass of the
initial [final] hyperon, $q = p_1 - p_2$ is the four-momentum transfer and
$V_{\rm CKM}$ stands for either $V_{ud}$ or $V_{us}$. The quantities $f_1$
and $g_1$ are the vector and axial-vector form factors, $f_2$ and $g_2$
are the weak magnetism and electricity form factors, while $f_3$ and $g_3$
are the induced scalar and pseudoscalar form factors, respectively. Time
reversal invariance requires that the form factors be real. The six form
factors are functions of $q^2$ and, unless explicitly noted otherwise,
their values at $q^2 =0$ are discussed. $f_3$ and $g_3$ may be safely
ignored in decays to an electron, because their contributions to the
different observables are suppressed by the electron mass.

In the limit of exact flavor $SU(3)$ symmetry, the hadronic weak vector
and axial-vector currents belong to $SU(3)$ octets, so  the form factors
of different HSD are related by $SU(3)$ flavor symmetry,
\begin{eqnarray}
f_k (q^2) & = & C_F^{B^\prime B} F_k(q^2) + C_D^{B^\prime B} D_k(q^2) \, ,
\label{eq:fone} \\
g_k (q^2) & = & C_F^{B^\prime B} F_{k+3}(q^2) + C_D^{B^\prime B} 
D_{k+3}(q^2) \, ,
\label{eq:gone}
\end{eqnarray}
where $F_i(q^2)$ and $D_i(q^2)$ are reduced form factors and
$C_F^{B^\prime B}$ and $C_D^{B^\prime B}$ are well-known Clebsch-Gordan
coefficients. The weak currents and the electromagnetic current are
members of the same $SU(3)$ octet, so all the vector form factors for HSD
are related at $q^2 = 0$ to the electric charges and the anomalous
magnetic moments of the nucleons $\kappa_{p,n}$. In particular, $F_1(0) =
1$, $D_1(0) = 0$, $F_2(0) = \kappa_p + \frac12 \kappa_n$, $D_2(0)
=-\frac32 \kappa_n$. Additionally, the conservation of the electromagnetic
current implies $F_3(q^2) = D_3(q^2) = 0$ so that the form factor
$f_3(q^2)$ vanishes for all HSD in the $SU(3)$ symmetry limit.

The leading axial-vector $g_1$ form factor is given in terms of two
reduced form factors, $D$ and $F$. The $g_2$ form factor for diagonal
matrix elements of hermitian currents (e.g. $\langle B | \bar u \gamma^\mu
\gamma_5 u - \bar d \gamma^\mu \gamma_5 d| B \rangle $) vanishes by
hermiticity and time-reversal invariance. $SU(3)$ symmetry then implies
that $g_2=0$ in the symmetry limit.

For the decuplet baryons, we will follow a formalism consistent with
chiral symmetry adopted in Ref.~\cite{dai} and originally introduced by
Peccei~\cite{pec}. In this formalism, the width of a decuplet baryon
$B^\prime$ decaying to an octet baryon $B$ is given by
\begin{eqnarray}\label{peccei}
\Gamma_{B^\prime} = \frac{g^2 {C(B,B^\prime)}^2(E_B + M_B) {\rm q_\pi^3}}
{24 \pi f_\pi^2 M_{B^\prime}} \, ,
\end{eqnarray}
where $E_B$ and ${\rm q_\pi}$ are the octet baryon energy and the pion
three-momentum in the rest frame of $B^\prime$, $f_\pi = 93$ MeV is the
pion decay constant, $g$ is the axial-vector coupling and $C(B,B^\prime)$
is a Clebsch-Gordan coefficient [$C(B,B^\prime)$ = 1, $1/\sqrt{2}$,
$1/\sqrt{3}$, $1/\sqrt{2}$ for $\Delta \rightarrow N \pi$, $\Sigma^*
\rightarrow \Lambda \pi$, $\Sigma^* \rightarrow \Sigma \pi$, $\Xi^*
\rightarrow \Xi \pi$, respectively].

\section{OPERATOR ANALYSIS}\label{sec:OA}

In the $N_c \to \infty$ limit, it has been shown that the baryon sector
has a contracted $SU(2F)$ spin-flavor symmetry, where $F$ is the number of
light quark flavors~\cite{dm,gsk}. Corrections to the large $N_c$ limit
can be expressed in terms of $1/N_c$-suppressed operators with
well-defined spin-flavor transformation properties~\cite{dm}.  Recently,
the $1/N_c$ expansion has yielded predictions for properties of baryons
such as axial-vector couplings and magnetic
moments~\cite{djm,dashen96,dai}  which are in good agreement with the
experimental data.  The $1/N_c$ expansion of QCD using quark operators as
the operator basis~\cite{cgo,lm,dashen96} provides a framework for
studying the spin-flavor structure of baryons. In the case of three
flavors, the lowest lying baryon states fall into a representation of the
spin-flavor group $SU(6)$. When $N_c = 3$, this corresponds to the very
well-known {\bf 56} dimensional representation of $SU(6)$.

A complete set of operators can be constructed using the zero-body
operator $\openone$ and the one-body operators
\begin{eqnarray}
J^i & = & q^\dagger \left(\frac{\sigma^i}{2} \otimes \openone \right) q
\qquad (1, 1) \, , \nonumber \\
T^a & = & q^\dagger \left( \openone \otimes \frac{\lambda^a}{2} \right) q
\qquad (0, 8) \, , \label{eq:gen} \\
G^{ia} & = & q^\dagger \left( \frac{\sigma^i}{2} \otimes
\frac{\lambda^a}{2} \right) q \qquad (1, 8) \, , \nonumber
\end{eqnarray}
where $J^i$ are the baryon spin generators, $T^a$ are the baryon flavor
generators, and $G^{ia}$ are the baryon spin-flavor generators. The
transformation properties of these generators under $SU(2)\times SU(3)$
are given explicitly in Eq.~(\ref{eq:gen}) as $(j,d)$, where $j$ is the
spin and $d$ is the dimension of the $SU(3)$ flavor representation.

Any QCD one-body operator transforming according to a given $SU(2)\times
SU(3)$ representation has a $1/N_c$ expansion of the form
\begin{eqnarray}
{\cal O}_{\rm QCD} = \sum_{n=0}^{N_c} c_n \frac{1}{N_c^{n-1}}{\cal O}_n 
\label{eq:genex} \, ,
\end{eqnarray}
where $c_n(1/N_c)$ are unknown coefficients which have power series
expansions in $1/N_c$ beginning at order unity. The sum in
Eq.~(\ref{eq:genex}) is over all possible independent $n$-body operators
${\cal O}_n$ with the same spin and flavor quantum numbers as ${\cal
O}_{\rm QCD}$. The use of operator identities~\cite{dashen96} reduces the
operator basis to independent operators. In this analysis we are concerned
with the $1/N_c$ expansions of the QCD vector and axial vector currents,
whose matrix elements between $SU(6)$ symmetric states give the HSD form
factors.

The $1/N_c$ expansion for the HSD amplitudes is derived to first order in
flavor symmetry breaking, and to  leading order in $1/N_c$ for most of the
form factors.  For the $f_1$ form factor, however, we include second-order
flavor symmetry breaking corrections, since the Ademollo-Gatto theorem
states that there are no first order corrections, so that the leading
symmetry breaking correction to $f_1$ is of second order. A chiral
perturbation theory calculation shows that (formally) second order
symmetry breaking effects actually contribute at first order in symmetry
breaking~\cite{kra,and},\footnote{For a more detailed explanation of this
seemingly contradictory statement, see Ref.~\cite{and}.} so we have
included these effects. The $f_2$ form factor is multiplied by $q$, and so
makes a small contribution to the HSD amplitude. Since $q$ is of order the
hyperon mass differences, the contribution of the first order $SU(3)$
symmetry breaking correction in $f_2$ to the HSD amplitude is comparable
to a second order symmetry breaking effect, and is neglected.  In the
symmetry limit $f_2$ can be determined from the baryon anomalous magnetic
moments, and that is what we do here. The axial form factor $g_1$ is
computed to first order in symmetry breaking. The $g_2$ form factor
vanishes in the symmetry limit, so its contribution is comparable to
symmetry breaking terms in $f_2$, and is neglected. Finally, $f_3$ and
$g_3$ contributions are proportional to the electron mass, and also will
be neglected.

\subsection{Vector form factor $f_1$}

We begin by deriving the $1/N_c$ expansion for the baryon vector current
in the $SU(3)$ flavor symmetry limit. At $q^2 = 0$, the hyperon matrix
elements for the vector current are given by the matrix elements of the
associated charge or $SU(3)$ generator. Let $V^{0a}$ denote the flavor
octet baryon charge\footnote{The subscript QCD emphasizes the fact that
$\overline {q}$ and $q$ are QCD quark fields, not the quark creation and
annihilation operators of the quark representation.}
\begin{eqnarray}
V^{0a} = \left< B^\prime \left|{\left(\overline {q} \gamma^0
\frac{\lambda^a}{2}q\right)}_{\rm QCD} \right| B \right> \, ,
\end{eqnarray}
whose matrix elements between $SU(6)$ symmetric states give the values of
the leading vector form factor $f_1$. $V^{0a}$ is spin-0 and a flavor
octet, so it transforms as (0,8) under $SU(2)\times SU(3)$.

The $1/N_c$ expansion for a (0,8) operator was obtained in Ref.~\cite{jl}.
Operator reduction rules imply that only $n$-body operators with a single
factor of either $T^a$ or $G^{ia}$ appear. Thus, the allowed one- and
two-body operators are
\begin{eqnarray}
&   & {\cal O}_1^a = T^a \, , \\
&   & {\cal O}_2^a = \{J^i,G^{ia}\} \, .
\end{eqnarray}
The remaining operators are obtained from these operators by
anticommuting with $J^2$, ${\cal O}_{n+2} = \{J^2, {\cal O}_n\}$. Thus,
the $1/N_c$ expansion of $V^{0a}$ has the form
\begin{eqnarray}
V^{0a} = \sum_{n=1}^{N_c} c_n \frac{1}{N_c^{n-1}}{\cal O}_n^a 
\label{eq:vcoup} \, .
\end{eqnarray}
The operator $V^{0a}$ at $q^2=0$ is a special (0,8) operator; it is the
generator of $SU(3)$ symmetry transformations. This fixes
\begin{eqnarray}
c_1 = 1, \qquad c_n = 0,\ n > 1.
\end{eqnarray}
Thus, the $1/N_c$ expansion of $V^{0a}$ in the limit of exact $SU(3)$
flavor symmetry reduces to
\begin{eqnarray}
V^{0a} = T^a \, , \label{eq:vafin}
\end{eqnarray}
to all orders in the $1/N_c$ expansion. The matrix elements of 
Eq.~(\ref{eq:vafin}) will be denoted by $f_1^{\rm SU(3)}$ hereafter. 

\subsection{Vector form factor with perturbative $SU(3)$ breaking}

In QCD, flavor $SU(3)$ symmetry breaking is due to the strange quark mass
$m_s$, and transforms as a flavor octet. In order to construct the most
general $1/N_c$ expansion for $V^{0a}$ up to second-order in symmetry
breaking, we need to consider the spin-0 $SU(2)\times SU(3)$
representations of the quark operators contained in the $SU(6)$
representations {\bf 1}, {\bf 35}, {\bf 405} and {\bf 2695}, {\it i.e.}
$(0,1)$, $(0,8)$, $(0,27)$, $(0,64)$, and $(0,10+\overline {10})$, since
the baryon $1/N_c$ expansion extends only to three-body operators if we
restrict ourselves to physical baryon states.\footnote{The $(0,10-
\overline {10})$ representation is not allowed by time reversal 
invariance.} The $1/N_c$ expansions for the above representations were
computed in Ref.~\cite{jl}; the results can be summarized as follows.

The $1/N_c$ expansion for a $(0,1)$ QCD operator starts with the zero-body
operator ${\cal O}_0 = \openone$. Additional operators are obtained by
anticommuting with $J^2$.

The $1/N_c$ expansion for a $(0,8)$ operator has the same form  as
Eq.~(\ref{eq:vcoup}) and will not be repeated here. The $1/N_c$ expansion
for a  $(0,27)$ operator contains the two- and three-body operators
\begin{eqnarray}
&   & {\cal O}_2^{ab} = \{T^a, T^b\} \, , \\
&   & {\cal O}_3^{ab} = \{T^a, \{J^i, G^{ib}\}\} + \{T^b, \{J^i,
G^{ia}\}\} \, ,
\end{eqnarray}
where the flavor singlet and octet components of the above operators have
to be subtracted off. As for a $(0,64)$ operator, the $1/N_c$ expansion
starts with a single three-body operator
\begin{eqnarray}
{\cal O}_3^{abc} = \{T^a, \{T^b, T^c \}\} \, ,
\end{eqnarray}
where it is understood that the singlet, octet and 27 components are
subtracted off in such a way that only the 64 component remains. Finally,
for a $(0,10+\overline{10})$ operator, one obtains
\begin{eqnarray}
{\cal O}_3^{ab} = \{T^a, \{J^i,G^{ib}\}\} - \{T^b,\{J^i,G^{ia}\}\} \, .
\end{eqnarray}

First-order symmetry breaking terms in $V^{0a}$ are given by setting one
free flavor index equal to 8 in the operators described above. At
second-order in the symmetry breaking, two free flavor indices are set
equal to 8. This gives
\begin{eqnarray}
V^{0a} + \delta V^{0a} & = & (1 + \epsilon a_1)  T^a  + 
\epsilon a_2 \frac{1}{N_c} \{J^i,G^{ia}\} +
\epsilon a_3 \frac{1}{N_c^2} \{J^2,T^a\} \nonumber \\
&&+\epsilon b_1 d^{ab8} T^b + \epsilon
b_2 \frac{1}{N_c} d^{ab8}\{J^i,G^{ib}\} \nonumber \\
&& + \epsilon b_3 \frac{1}{N_c^2} d^{ab8}\{J^2,T^b\} +
\epsilon a_4\frac{1}{N_c} \{T^a,T^8\} \nonumber \\
&   & \mbox{} + \epsilon a_5 \frac{1}{N_c^2} \left( \{T^a, \{J^i,
G^{i8}\}\} + \{T^8, \{J^i,G^{ia}\}\} \right) \nonumber \\
&   & \mbox{} + \epsilon a_6 \frac{1}{N_c^2} \left( \{T^a, \{J^i,
G^{i8}\}\} - \{T^8, \{J^i,G^{ia}\}\} \right) \nonumber \\
&& + \epsilon^2 b_4 \frac{1}{N_c} d^{ab8}\{T^b,T^8\} +
\epsilon^2 a_7 \frac{1}{N_c^2} \{T^a,\{T^8,T^8\}\} \nonumber \\
&& + \epsilon^2 b_5 \frac{1}{N_c^2} d^{ab8} \left( \{T^b,
\{J^i, G^{i8}\}\} + \{T^8, \{J^i,G^{ib}\}\} \right) \nonumber \\
&   & \mbox{} + \epsilon^2 b_{6} \frac{1}{N_c^2} d^{ab8} \left( \{T^b,
\{J^i, G^{i8}\}\} - \{T^8, \{J^i,G^{ib}\}\} \right) \, ,
\label{eq:vazero}
\end{eqnarray}
where $\epsilon \sim m_s$ is a (dimensionless) measure of $SU(3)$
breaking. Observe that similar terms with the $d$-symbol replaced by an
$f$-symbol are ruled out by time reversal invariance. None of the (0,1)
operators contributes to $V^{0a}$ for $\Delta S=0$ and $|\Delta S|=1$ weak
decays, so they have been omitted in Eq.~(\ref{eq:vazero}). Note that the
coefficients in Eq.~(\ref{eq:vazero}) must be such that there is no
symmetry breaking for the $\Delta S=0$ weak decays, since isospin symmetry
is not broken by the strange quark mass.

Equation~(\ref{eq:vazero}) can be rewritten in terms of the number of
strange quarks, $N_s$, and the strange quark spin, $J_s^i$,
using~\cite{dashen96}
\begin{eqnarray}
&   & T^8 = \frac{1}{2\sqrt 3} (N_c - 3 N_s) \, , \label{eq:teight} \\
&   & G^{i8} = \frac{1}{2\sqrt 3}(J^i - 3J_s^i) \, . \label{eq:geight}
\end{eqnarray}
After making use of the identity~\cite{dai}
\begin{eqnarray}
J^i G^{ia}/J^2 = \frac23 \left( T^a + \frac12 \{T^a, N_s \} \right) \, ,
\end{eqnarray}
valid for $\Delta S = 2\Delta I$ transitions, rearranging terms and
absorbing factors of $N_c^{-1}$ and $N_c^{-2}$ we obtain a rather compact
form for $V^{0a}$, namely,
\[
V^{0a} = T^a,
\]
for $\Delta S=0$ decays, and 
\begin{eqnarray}
V^{0a} = (1 + v_1^\prime)T^a + v_2^\prime \{T^a, N_s\} + v_3^\prime \{T^a,
N_s^2\} + v_4^\prime \{T^a, J^2\} + v_5^\prime \{T^a, -I^2+J_s^2\}\, ,
\label{eq:vlast1}
\end{eqnarray}
for $|\Delta S| = 1$ decays. Here $I$ is the isospin. For the decays we are
considering, the $v_3^\prime$ and $v_4^\prime$ terms are not independent,  and
can be written as linear combinations of the $v_1^\prime$ and $v_2^\prime$
terms. Thus, Eq.~(\ref{eq:vlast1})  reduces to
\begin{eqnarray}
V^{0a} = (1 + v_1)T^a + v_2 \{T^a, N_s\} + v_3 \{T^a,-I^2+J_s^2\}\, .
\label{eq:vlast}
\end{eqnarray}
The baryons are eigenstates of $J^2$, $I^2$, $J_s^2$, and $N_s$, so the
matrix elements of Eq.~(\ref{eq:vlast}) can be computed straightforwardly.
They are listed in Table~\ref{t01} for the processes we are concerned
with.

\subsection{Axial-vector form factor $g_1$ }

The $1/N_c$ expansion for the axial-vector current $A^{ia}$ was  discussed
in great detail in Ref.~\cite{dashen96} and we will only state the answer
here. The axial current matrix elements can be written as
\begin{eqnarray}
\frac12 A^{ia} & = & a G^{ia} + bJ^i T^a + \Delta^a(c_1 G^{ia} +
c_2 J^iT^a) \nonumber \\
&   & \mbox{} + c_3 \{G^{ia},N_s\} + c_4\{T^a, J_s^i\} +
\frac{1}{\sqrt 3} \delta^{a8}W^i \nonumber \\
&   & \mbox{} - \frac{d}{2} ( \{J^2, G^{ia}\} - \frac12 \{J^i,
\{J^j,G^{ja}\}\} ) \, , \label{eq:vafit}
\end{eqnarray}
where
\begin{eqnarray}
W^i & = & (c_4 - 2c_1)J_s^i + (c_3 - 2c_2)N_s J^i - 3(c_3 + c_4)N_s J_s^i
\, ,
\end{eqnarray}
and $\Delta_a = 1$ for $a = 4,5,6$, or 7 and equals zero otherwise. A term
proportional to $d$ has been added in order for the $SU(3)$ symmetric
parameters $D$, $F$, and ${\cal C}$~\cite{jm} to have arbitrary values.
Adding this term will avoid the mixing between symmetry breaking effects
and $1/N_c$ corrections in the symmetric couplings. Furthermore, the
couplings have been parametrized in such a way that only the parameters
$a$, $b$, and $d$ contribute to processes which take place in the
strangeness-zero sector. Including $SU(3)$ breaking, the reduced form
factors $D$ and $F$ are defined as
\begin{equation}
D  =  a , \qquad
F  =  \frac23 a + b \, ,
\end{equation}
so that $g_1/f_1 = D + F$ is positive for neutron decay, which fixes all other
signs. Thus, for any process, the matrix elements of $A^{ia}$ are given as the
sum of the parameters $a$, $b$, $d$, $c_1, \ldots, c_4$ times matrix elements
of the operators involved in the expansion Eq.~(\ref{eq:vafit}). The operator
matrix elements were computed in Ref.~\cite{dai} and are listed in
Table~\ref{t02} for the sake of completeness.

\subsection{Weak magnetism form factor $f_2$}

In the limit of exact $SU(3)$ flavor symmetry, the weak magnetism form
factors $f_2$ are directly related to the anomalous magnetic moments of
the nucleons, and are given in terms of two invariants $m_1$ and $m_2$.
Since the magnetic moment is a spin-$1$ octet operator, it has a $1/N_c$
expansion identical in structure to the axial current. It is convenient to
define the two parameters $m_1$ and $m_2$ by
\[
M^i = m_1 G^{iQ} + m_2 J^i T^Q,
\]
where $Q$ represents the $SU(3)$ generator  which is the electric charge,
so $G^{iQ} \equiv G^{i3}+G^{i8}/\sqrt 3$, and $T^Q \equiv T^3 + T^8/\sqrt
3$. The parameters $m_{1,2}$ can be determined from the anomalous magnetic
moments of the hyperons.

The contributions of $f_2$ to the different observables of HSD in the
$SU(3)$ limit are first-order symmetry breaking contributions because of
the kinematic factor of $q$.  Previous work~\cite{rfm,gar92} has shown
that reasonable shifts from the $SU(3)$ predictions of $f_2$ do not have
any observable effect upon $\chi^2$ or $g_1$ in a global fit to
experimental data. We will use the best fit values~\cite{dai} $m_1=2.87$
and $m_2=-0.077$ obtained from the baryon anomalous magnetic moments to
fix $f_2$.

\subsection{Weak electricity form factor $g_2$}

In the $SU(3)$ flavor symmetry limit, the form factor $g_2$ vanishes, so that
$g_2$ is proportional to $SU(3)$ symmetry breaking at leading order. $g_2$
transforms oppositely to $g_1$ and $f_2$ under time-reversal, and therefore has
a different $1/N_c$ operator expansion. 

Let $W^{ia}$ be the operator whose matrix elements give the values  of $g_2$.
At first order in $SU(3)$ symmetry breaking, the contribution to $g_2$
transforms as $(1,8)$ and  $(1,10-\overline {10})$ under spin and flavor. The
$(1,8)$ expansion is given by
\begin{eqnarray}
\delta W_8^{ia} \propto i b_1 f^{ab8} G^{ib} +  i b_2 f^{ab8}\frac{J^iT^b}{N_c} \, ,
\label{eq:dw}
\end{eqnarray}
which involves the $f$-symbol, rather than the $d$-symbol since $g_2$ has the
opposite time-reversal properties from $g_1$ and $f_2$.  The
$(1,10-\overline{10})$ expansion has not been presented previously in the
literature. The operator $\{G^{ig},T^h\}-\{G^{ih},T^g\}$, which contains
$(10+\overline{10})$, can be split into 10 and $\overline{10}$ representations
by contracting with $f^{acg} d^{bch}$~\cite{dashen96}. The resulting operator
contains $i(10-\overline {10})$, which is $T$-even. This procedure leads to the
contribution
\begin{eqnarray}
\delta W_{10-\overline{10}}^{ia} \propto i f^{8cg}d^{ach} \left( \{
G^{ig},T^h \} - \{ T^g, G^{ih} \} \right) = - i f^{acg}d^{8ch} \left( \{
G^{ig},T^h \} - \{ T^g, G^{ih} \} \right) \, . \label{eq:dwb}
\end{eqnarray}
For HSD, the three operators Eq.~(\ref{eq:dw})-(\ref{eq:dwb}) are linear
combinations of the three allowed invariants
\begin{equation}\label{eq:dddw}
{\rm Tr\, } \left[T^a,T^8\right]\bar B B, \ {\rm Tr\, } \bar B \left[T^a,T^8\right] B,\
{\rm Tr\, } \bar B T^a  {\rm Tr\, } B T^8 - 
{\rm Tr\, } \bar B T^8 {\rm Tr\, } B T^a
\end{equation}
given by a general $SU(3)$ analysis~\cite{garcia85} neglecting isospin 
breaking. We choose as independent parameters $b_1$ and $b_2$ in
Eq.~(\ref{eq:dw}), and $b_3$ that multiplies the third invariant in
Eq.~(\ref{eq:dddw}). The matrix elements  are listed in Table~\ref{t04}. For
any process, the matrix elements of $W^{ia}$ can be given as a sum of the
parameters $b_{1-3}$ times the operator matrix elements listed in
Table~\ref{t04}.

The parameters $b_{1-3}$ are proportional to $\epsilon$. As mentioned in the
introductory remarks to this section, $g_2$ should be neglected for a
consistent analysis. Nevertheless, we tried to see if we could obtain some
information on $g_2$ from the experimental data using the above formul\ae\ for
$g_2$.  However, the data are not accurate enough for an extraction of the
small $g_2$-dependence of the decay amplitudes. One expects that $g_2$ should
be about 25\% of $f_2$, so that $g_2$ is $\lower0.3em \hbox{$\stackrel{<}
{\sim}$} 0.5$.

\section{FITTING THE DATA}\label{sec:FIT}

The experimentally measured quantities~\cite{prd} in HSD are the total decay
rate $R$, angular correlation coefficients  $\alpha_{e\nu}$, and angular
spin-asymmetry coefficients $\alpha_{e}$, $\alpha_\nu$, $\alpha_B$, $A$, and
$B$. Often, the data is presented in terms of $R$ and the ratio  $g_1/f_1$ for
the decay. This information is displayed in Tables~\ref{t05} and~\ref{t06} for
the measured decays. The theoretical expressions for the total decay rates and
angular coefficients can be found in Ref.~\cite{garcia85}. The radiative
corrections and the four-momentum-transfer contribution to the form factors are
also discussed in this reference. In the present analysis, we will take these
corrections into account.\footnote{In this work we have adopted a dipole form
for the leading vector and axial-vector form  factors, with masses $M_V = 0.84$
GeV and $M_A = 0.96$ GeV for $\Delta S = 0$ transitions, and $M_V = 0.97$ GeV
and $M_A = 1.11$ GeV for $|\Delta S| = 1$ processes~\cite{garcia85}.  In
Ref.~\cite{dai} somewhat different values of $M_V$ and $M_A$ were used. Our
results are insensitive to this difference.} The experimentally measured
quantity for the decuplet baryons is the decay width, which has been converted
to an axial-vector coupling for each decay using the Goldberger-Treiman
relation and Eq.~(\ref{peccei}). This information is displayed in
Table~\ref{t07}.

In this section we perform a number of different fits to the experimental data.
The experimental data  which are used are the decay rates and  the spin and
angular correlation coefficients. The value of $g_1/f_1$ is not included, since
it is determined from the other quantities and is not an independent
measurement. For $\Xi^- \to \Sigma^0$ decay, we have used $g_1/f_1$, however,
since the spin and angular correlation coefficients have not been measured. The
parameters to be fitted are those arising from the $1/N_c$ expansions for the
couplings, namely, $v_{1-3}$ for $f_1$ introduced in Eq.~(\ref{eq:vlast}) and
$a$, $b$, $d$, $c_{1-4}$ for $g_1$ given in Eq.~(\ref{eq:vafit}). We also
attempted to fit $b_{1-3}$ for $g_2$, but the experimental data is not
sufficiently accurate to determine the small $g_2$ contribution to the decay
amplitude. We therefore neglect $g_2$ in the rest of the analysis. Finally, in
the first stage of the analysis we use as inputs the PDG values of $V_{ud}$ and
$V_{us}$~\cite{prd} (which are primarily obtained from nuclear $\beta$ decay
and $K_{e3}$ decay, respectively). We later proceed to fit for them as well.

\subsection{SU(3) fit}

The simplest possible fit is an $SU(3)$ symmetric fit to HSD (ignoring the
decuplet decays) which involves only two parameters $a$, $b$ for $g_1$;
it corresponds to a fit using only $F$ and $D$. The results are $a = 0.797
\pm 0.006$, $b = -0.059 \pm 0.009$, which yield $F = 0.47 \pm 0.01$, $D =
0.80 \pm 0.01$, $3F-D = 0.62 \pm 0.03$,  with $\chi^2 = 62.3$ for 23
degrees of freedom. The large $\chi^2$ of the fit is clear evidence for
$SU(3)$ breaking. A similar fit using the rates and $g_1/f_1$ ratios was
performed in  Ref.~\cite{dai}. Both results are in very good agreement. We
also followed this reference in order to make a preliminary study of
$\Delta S=0$ decays only. Our fits produce similar results and there is no
need to show them here.

\subsection{First-order symmetry breaking}

The next step is to see how the results are modified once first-order symmetry
breaking is taken into account. To this order, $f_2$ will be kept at its
$SU(3)$ symmetric value and $g_2$ is set to zero. $f_1$ is also kept at its
symmetry-limit value, $f_1^{\rm SU(3)}$, because of the Ademollo-Gatto
theorem.\footnote{ We have mentioned earlier that chiral corrections to $f_1$
are not necessarily second order. They are included in the next section.} Thus,
only the order $\epsilon$ terms in $g_1$ introduced in Eq.~(\ref{eq:vafit})
will be considered. Fitting $a$, $b$, $d$, $c_{1-4}$ leads to the results
listed as Fit~A of Table~\ref{t09}, with $\chi^2 = 51.6$ for 22 degrees of
freedom. Comparing the values of $F$ and $D$ with those of the previous
section, we see that the change due to symmetry breaking corrections is in fact
small (compared to a naive estimate of $SU(3)$ breaking). The leading parameter
$a$ is order  unity, $b$ is order $1/N_c$, $d$ is order $1/N_c^2$, and the
values of $c_{1-4}$ are small or smaller than expected from first-order
symmetry breaking ($\epsilon \sim 30\%$,  which is a measure of symmetry
breaking) and factors of $1/N_c$.  These results agree with the ones presented
in Ref.~\cite{dai}, which were obtained by using  the total decay rates and
$g_1/f_1$ ratios as experimental inputs.

Notice that the quantity $3F - D$, which is relevant for the analysis of
spin-dependent deep inelastic scattering, is smaller than its value determined
in the $SU(3)$ limit, and is considerably smaller than its $SU(6)$ symmetric
value of 1~\cite{ehr,lich,dai,sav}. Before drawing any conclusion, however, we
will study in the next section the effect of symmetry breaking in $f_1$ on the
different observables, and in particular upon the reduced form factors $F$ and
$D$. 

\subsection{Symmetry breaking in $f_1$}

In the previous sections $f_1$ was fixed at its $SU(3)$ symmetric value,
$f_1^{\rm SU(3)}$. We now proceed to incorporate symmetry breaking
corrections into the $f_1$ form factors in $|\Delta S| = 1$  decays.
Formally, one expects that these corrections should be second-order in
symmetry breaking, due to the Ademollo-Gatto theorem. However, we know
from explicit computations of chiral loops~\cite{kra,and} that there are,
in fact, corrections which can be considered to be first order in symmetry
breaking. These were not included in Ref.~\cite{dai}.

The best fit parameters $v_{1-3}$ for $f_1$ and $a$, $b$, $d$, $c_{1-4}$, for
$g_1$ are displayed as Fit~B in  Table~\ref{t09}. The resulting form factors
are given in Tables~\ref{t10}  and~\ref{t11}. The theoretical predictions for
the different observables are  listed in Tables~\ref{t12}, \ref{t13}
and~\ref{t14} for the sake of completeness.  The fit has $\chi^2 = 39.2$ with
19 degrees of freedom.

{}From Table~\ref{t11}, we observe that SU(3) breaking corrections to the 
leading vector form factors $f_1$ are as much as  12\%, depending on the
strange-quark content of the decaying and emitted baryons. Furthermore, we
can observe that the natural trend is $f_1/f_1^{\rm SU(3)} > 1$, as was
pointed out in  Refs.~\cite{rfm,gar92}. Additionally, the ratios $g_1/f_1$
of Fit B (in Tables~\ref{t13} and \ref{t14}) agree with the experimental
ones listed in Tables~\ref{t05}  and~\ref{t06}. As for the axial-vector
couplings of the decuplet baryons, we can see in Table~\ref{t12} that the
theoretical predictions are in good agreement with their experimental
values. The highest contribution to $\chi^2$ comes from  $\Sigma^*
\rightarrow \Lambda \pi$ decay.

In Tables~\ref{t13} and~\ref{t14}, the predictions for the different
observables are in reasonable agreement with their experimental counterparts
displayed in Tables~\ref{t05} and~\ref{t06}, respectively. The highest
contributions to the total $\chi^2$ arise mainly from $\alpha_e$ ($\Delta
\chi^2 = 2.7$) in $n \rightarrow p e^- \overline {\nu}_e$, $\alpha_e$ ($\Delta
\chi^2 = 2.4$), $\alpha_\nu$ ($\Delta \chi^2 = 6.6$) in $\Lambda \rightarrow p
e^- \overline{\nu}_e$, $\alpha_\nu$ ($\Delta \chi^2 = 3.8$) in $\Sigma^-
\rightarrow n  e^- \overline{\nu}_e$, and $R$ ($\Delta \chi^2=2.2$)  and
$g_1/f_1$ ($\Delta \chi^2 = 5.7$) in $\Xi^- \rightarrow \Sigma^0 e^-
\overline{\nu}_e$.  If $\alpha_\nu$ in $\Lambda \to p e^- \overline{\nu}_e $
decay is left out, there are small readjustments of the parameters and
predicted observables, some of them almost imperceptible, so that one can draw
the same conclusions as above. This fact suggests that there is an experimental
inconsistency in the value of this $\alpha_\nu$.

One can redo the above fit including $g_2$. The best fit parameters are
$b_1=0.9 \pm 0.5$, $b_2=1.0 \pm 0.2$, $b_3=-0.0 \pm 0.4$, and $\chi^2$ is
reduced to 27 for 16 degrees of freedom. The reduction in $\chi^2$ suggests
that there is a non-zero $g_2$, but the large error bars indicate that the
experimental data is not sufficiently accurate to determine 
$g_2$.\footnote{For example, fitting to all the experimental data except
$g_1/f_1$ in $\Xi^- \to \Sigma^0$ decay gives completely different values: 
$b_1=-0.7 \pm 1.2$, $b_2=0.6  \pm 0.4$, $b_3=0.0 \pm 0.4$.} The fit including
$g_2$ gives $3F-D=0.3\pm 0.1$, and $f_1=1.4\pm0.1$, $g_1=1.4\pm 0.1 $ for
$\Xi^0 \to \Sigma^+$ decay.

Finally, we fit the data with the CKM matrix elements $V_{ud}$ and $V_{us}$ as
free parameters (neglecting $g_2$).  Unfortunately, there is not enough
information on the $|\Delta S| = 1$ decays to make a detailed analysis and
extract a value of $V_{us}$ from these data only. We will content ourselves
with performing a global fit to data allowing both $V_{ud}$ and $V_{us}$ to be
free parameters. The best fit values for the CKM parameters are
\begin{equation}
V_{ud}  =  0.9743 \pm 0.0009,\qquad V_{us}  =  0.2194 \pm 0.0023 \,.
\end{equation}
These values have to be compared to their PDG counterparts~\cite{prd} which are
$V_{ud} = 0.9736 \pm 0.0010$ and $V_{us} = 0.2196 \pm 0.0023$. The latter is
the one quoted from $K_{e3}$ decays. The best fit values for the other
parameters are listed as Fit~C in the tables. The values for the HSD parameters
in Fit~C are indistinguishable from Fit~B, and have not been listed separately
in Tables~\ref{t13} and \ref{t14}.

\subsection{Errors}

The fits to the experimental data have used theoretical expressions for $SU(3)$
breaking in the $f_1$ and $g_1$ form factors at leading order in the $1/N_c$
expansion. The theoretical errors are of order $\epsilon/N_c$, where $\epsilon$
is a measure of $SU(3)$ breaking. In most hadronic quantities, $SU(3)$ breaking
is of order 20--30\%, so $\epsilon/N_c$ is of order 5--10\%. One can also get a
measure of the uncertainty in the results from the fit itself. One can use the
PDG procedure~\cite{prd} for rescaled errors to reduce $\chi^2$ to one per
degree of freedom. This multiplies all the errors on Fit C by $1.4$. It is
important to keep in mind that the tables list only the errors obtained from
$\chi^2$ fits to the data, and do not include theory errors or any rescaling
factors. 

The quantity $3F-D$ is not well-determined. Small changes in the fit tend to
move $F$ and $D$ in opposite directions, so that there are large changes in
$3F-D$. As an example of a theoretical uncertainty, consider using
Eq.~(\ref{peccei}) for the decuplet widths with the factor $(E_B+M_B)/M_{B'}$
omitted. This modified formula is what is obtained~\cite{dob} if one computes
the decuplet decay widths using the baryon chiral perturbation theory formalism
of Ref.~\cite{jm}. The modification of Eq.~(\ref{peccei}) is equivalent to
changes of order $\epsilon/N_c$ in the theoretical formul\ae\ used. The best
fit value for $3F-D$ changes to 0.46, and for $g_1$ in $\Xi^0 \to \Sigma^+$
$\beta$-decay becomes 1.07. The fitted parameters are listed as Fit~D in
Table~\ref{t10}. The difference between these numbers and those in Fit~C can be
regarded as an estimate of the theoretical uncertainty in the fits.

One can redo the fits using a (fixed) non-zero value of $g_2$ with $b_{1-3}$ of
the estimated theoretical size of $\lower0.3em \hbox{$\stackrel{<} {\sim}$}
0.5$. This changes the value of $g_1$ in $\Xi^0 \to \Sigma^+$ by 5--10\%, which
is consistent with the estimated theoretical uncertainty.

\section{CONCLUSIONS}\label{sec:RESULTS}

In this work we have analyzed the pattern of SU(3) symmetry breaking in the HSD
form factors within the $1/N_c$ expansion. We have incorporated second-order
symmetry-breaking corrections to the leading vector form factor $f_1$; $f_2$
was kept at its value predicted by $SU(3)$ symmetry, and $g_2$ was kept at its
$SU(3)$ symmetry value of zero. Additionally, we have corrected the
axial-vector form factors $g_1$ to first order in symmetry breaking. In the
several different fits to the experimental data we found that symmetry breaking
corrections to $f_1$ increase their magnitudes over their $SU(3)$ symmetric
predictions by up to 12\%, and that corrections to $g_1$ are consistent with
expectations.

We can predict the form factors for $\Xi^0 \to \Sigma^+$ $\beta$-decay, which
will soon be measured by KTeV~\cite{swallow}. Isospin symmetry relates this
decay to $\Xi^- \to \Sigma^0$, $z(\Xi^0 \to \Sigma^+) = \sqrt 2 z(\Xi^- \to
\Sigma^0)$, where $z$ is any of the form factors $f_i$ or $g_i$. This can be
seen explicitly from the tables. The measurement
of $\Xi^0$ $\beta$-decay will provide some very important information on
$SU(3)$ breaking in HSD. An $SU(3)$ symmetric fit predicts that $g_1$ for
$\Xi^0 \to \Sigma^+$ decay is about $1.27$. Directly using the measured value
of $g_1/f_1$ for $\Xi^- \to \Sigma^0$ decay and the $SU(3)$ symmetry value for
$f_1$ predicts that $g_1$ for $\Xi^0$ decay should be $1.29 \pm 0.16$. The
$SU(3)$ breaking analysis of this paper predicts that $g_1$ should have a
smaller value, in the range $1.02$--$1.07$. This number was obtained from a
combined fit of HSD and pionic decays of the decuplet baryons, which are
related in the $1/N_c$ approach. The fit is not entirely satisfactory, and it
appears that some of the experimental inputs are not consistent. Nevertheless,
the result that $g_1$ for $\Xi^0$ decay (and also $3F-D$) is smaller than its
$SU(3)$ symmetric value is robust. An $SU(3)$ breaking fit using only HSD data
(without including decuplet decays) would give a value for $g_1$ that is larger
than the $SU(3)$ symmetric value of $1.27$. As noted in Ref.~\cite{dai}, there
is clear evidence for $SU(3)$ breaking in the decuplet decays. At leading order
in the $1/N_c$ expansion, this necessarily implies $SU(3)$ breaking in the
hyperon $\beta$-decays, and leads to smaller values for $g_1$ in the $\Xi$
$\beta$-decays than an $SU(3)$ symmetric fit.

Before closing, let us stress the fact that the pattern of flavor symmetry
breaking lowers the values of $F/D$ and $3F-D$ with respect to their $SU(6)$
predictions of $2/3$ and $1$, respectively, as was observed previously in
Refs.~\cite{ehr,lich,dai,sav}. A further improvement on the parameters obtained
in the present work can come from additional or better measurements on the
several observables in HSD and decuplet baryons. However, with the current
available data, the $1/N_c$ expansion provides a reasonable framework to
analyze flavor SU(3) breaking in HSD in a model-independent way. One can regard
Fits~C and D as best fits to $SU(3)$ breaking with the current data, and the
difference between Fits C and D as an estimate of the theoretical uncertainty
in the results.

\acknowledgments
We would like to thank J.~Rosner and E.~Swallow for helpful discussions. We are
particularly grateful to E.~Swallow for detailed comments on the manuscript.
This work was supported in part by a Department of Energy grant
DOE-FG03-97ER40546. RFM was supported by a CONACYT Postdoctoral Fellowship
(Mexico) under the UC-CONACYT agreement of cooperation. EJ was supported in
part by the Alfred P. Sloan Foundation and by a National Young Investigator
award PHY-9457911 from the National Science Foundation.


\begin{table}
\squeezetable
\caption{Operator matrix elements for the vector form factor $f_1$.}
\label{t01}
\begin{tabular}{lddddd}
Transition & $f_1^{\rm SU(3)}$ & $v_1$ & $v_2$ & $v_3$ \\
\tableline
$n \rightarrow p$ & 1 & 0 & 0 & 0 \\
$\Sigma^\pm \rightarrow \Lambda$ & 0 & 0 & 0 & 0 \\
$\Lambda \rightarrow p$ & $-\sqrt{3/2}$ & $-\sqrt{3/2}$ & 
$-\sqrt{3/2}$ & 0 \\
$\Sigma^- \rightarrow n$ & $-$1 & $-$1 & $-$1 & $2$ \\
$\Xi^- \rightarrow \Lambda$ & $\sqrt{3/2}$ & $\sqrt{3/2}$ &
$3\sqrt{3/2}$ & $\sqrt{6}$ \\
$\Xi^- \rightarrow \Sigma^0$ & $1/\sqrt{2}$ & $1/\sqrt{2}$ &
$3/\sqrt{2}$ & 0 \\
$\Xi^0 \rightarrow \Sigma^+$ & 1 & 1 & 3 & 0
\end{tabular}
\end{table}


\begin{table}
\squeezetable
\caption{Operator matrix elements for the axial form factor $g_1$.}
\label{t02}
\begin{tabular}{lddddddd}
Transition & $a$ & $b$ & $d$ & $c_1$ & $c_2$ & $c_3$ & $c_4$ \\
\tableline
$\Delta \rightarrow N$ & $-2$ & 0 & 9/2 & 0 & 0 & 0 & 0 \\
$\Sigma^* \rightarrow \Lambda$ & $-2$ & 0 & 9/2 & 0 & 0 & $-4$ & 0 \\
$\Sigma^* \rightarrow \Sigma$ & $-2$ & 0 & 9/2 & 0 & 0 & $-4$ & 8 \\
$\Xi^* \rightarrow \Xi$ & $-2$ & 0 & 9/2 & 0 & 0 & $-8$ & 4 \\
$n \rightarrow p$ & 5/3 & 1 & 0 & 0 & 0 & 0 & 0 \\
$\Sigma^\pm \rightarrow \Lambda$ & $\sqrt{2/3}$ & 0 & 0 & 0 & 0 &
$\sqrt{8/3}$ & 0 \\
$\Lambda \rightarrow p$ & $-\sqrt{3/2}$ & $-\sqrt{3/2}$ & 0 &
$-\sqrt{3/2}$ & $-\sqrt{3/2}$ & $-\sqrt{3/2}$ & $-\sqrt{3/2}$ \\
$\Sigma^- \rightarrow n$ & 1/3 & $-$1 & 0 & 1/3 & $-$1 & 1/3 & 1/3 \\
$\Xi^- \rightarrow \Lambda$ & $1/\sqrt{6}$ & $\sqrt{3/2}$ & 0 &
$1/\sqrt{6} $ & $\sqrt{3/2}$ & $\sqrt{3/2}$ & 7/$\sqrt{6}$ \\
$\Xi^- \rightarrow \Sigma^0$ & $5/\sqrt{18}$ & $1/\sqrt{2}$ & 0 &
5/$\sqrt{18}$ & 1/$\sqrt{2}$ & 5/$\sqrt{2}$ & 1/$\sqrt{2}$ \\
$\Xi^0 \rightarrow \Sigma^+$ & 5/3 & 1 & 0 & 5/3 & 1 & 5 & 1
\end{tabular}
\end{table}


\begin{table}
\squeezetable
\caption{Operator matrix elements for the weak electricity form factor $g_2$.}
\label{t04}
\begin{tabular}{ldddd}
Transition & $b_1$ & $b_2$ & $b_3$  \\
\tableline
$n \rightarrow p$ & 0 & 0 & 0  \\
$\Sigma^\pm \rightarrow \Lambda$ & 0 & 0 & 1 \\
$\Lambda \rightarrow p$ & $-\sqrt{3/2}$ & $-\sqrt{3/2}$ & $-1$ \\
$\Sigma^- \rightarrow n$ & 1/3 & $-$1 & 0 \\
$\Xi^- \rightarrow \Lambda$ & $1/\sqrt{6}$ & $\sqrt{3/2}$ & 1 \\
$\Xi^- \rightarrow \Sigma^0$ & $5/\sqrt{18}$ & $1/\sqrt{2}$ & 0 \\
$\Xi^0 \rightarrow \Sigma^+$ & 5/3 & 1 & 0
\end{tabular}
\end{table}


\begin{table}
\squeezetable
\caption{Experimental data on three measured $\Delta S = 0$ hyperon 
semileptonic decays. The units of $R$ are 10$^{-3}$ s$^{-1}$ for neutron
decay and 10$^6$ s$^{-1}$ for the remaining decays.}
\label{t05}
\begin{tabular}{
l
r@{.}l@{\,$\pm$\,}r@{.}l r@{.}l@{\,$\pm$\,}r@{.}l r@{.}l@{\,$\pm$\,}r@{.}l
}
&
\multicolumn{4}{c}{$n \rightarrow p e^- \overline \nu_e$} &
\multicolumn{4}{c}{$\Sigma^+ \rightarrow \Lambda e^+ \nu_e$} &
\multicolumn{4}{c}{$\Sigma^- \rightarrow \Lambda e^- \overline \nu_e$} \\
\tableline
$R$ &
1 & 1274 & 0 & 0025 & 0 & 250  & 0 & 063 & 0 & 387  & 0 & 018 \\
$\alpha_{e\nu}$ &
$-$0 & 0766 & 0 & 0036 & $-$0 & 35 & 0 & 15 & $-$0 & 404 & 0 & 044 \\
$\alpha_e$ &
$-$0 & 08559 & 0 & 00086 & \multicolumn{4}{c}{} & \multicolumn{4}{c}{} \\
$\alpha_\nu$ &
0 & 990 & 0 & 008 & \multicolumn{4}{c}{} & \multicolumn{4}{c}{} \\
$A$ &
\multicolumn{4}{c}{} & \multicolumn{4}{c}{} & 0 & 07 & 0 & 07 \\
$B$ &
\multicolumn{4}{c}{} & \multicolumn{4}{c}{} & 0 & 85 & 0 & 07 \\
$g_1/f_1$ &
1 & 2601 & 0 & 0025 & \multicolumn{4}{c}{} & \multicolumn{4}{c}{}
\end{tabular}
\end{table}


\begin{table}
\squeezetable
\caption{Experimental data on four measured $|\Delta S| = 1$ hyperon 
semileptonic decays. The units of $R$ are 10$^6$ s$^{-1}$.}
\label{t06}
\begin{tabular}{
l
r@{.}l@{\,$\pm$\,}r@{.}l r@{.}l@{\,$\pm$\,}r@{.}l
r@{.}l@{\,$\pm$\,}r@{.}l r@{.}l@{\,$\pm$\,}r@{.}l
}
&
\multicolumn{4}{c}{$\Lambda \rightarrow p e^- \overline \nu_e$} &
\multicolumn{4}{c}{$\Sigma^- \rightarrow n e^-\overline \nu_e$} &
\multicolumn{4}{c}{$\Xi^- \rightarrow \Lambda e^- \overline \nu_e$} &
\multicolumn{4}{c}{$\Xi^-\rightarrow \Sigma^0 e^- \overline \nu_e$} \\
\tableline
$R$ &
3 & 161 & 0 & 058 & 6 & 876 & 0 & 235 & 3 & 435 & 0 & 192 &
0 & 531 & 0 & 104 \\
$\alpha_{e\nu}$ &
$-$0 & 019 & 0 & 013 & 0 & 347 & 0 & 024 & 0 & 53 & 0 & 10 &
\multicolumn{4}{c}{} \\
$\alpha_e$ &
0 & 125 & 0 & 066 & $-$0 & 519 & 0 & 104 & \multicolumn{4}{c}{} &
\multicolumn{4}{c}{} \\
$\alpha_\nu$ &
0 & 821 & 0 & 060 & $-$0 & 230 & 0 & 061 & \multicolumn{4}{c}{} &
\multicolumn{4}{c}{} \\
$\alpha_B$ &
$-$0 & 508 & 0 & 065 & 0 & 509 & 0 & 102 & \multicolumn{4}{c}{} &
\multicolumn{4}{c}{} \\
$A$ &
\multicolumn{4}{c}{} & \multicolumn{4}{c}{} & 0 & 62 & 0 & 10 &
\multicolumn{4}{c}{} \\
$g_1/f_1$ &
0 & 718 & 0 & 015 & $-$0 & 340 & 0 & 017 & 0 & 25 & 0 & 05 &
1 & 287 & 0 & 158
\end{tabular}
\end{table}


\begin{table}
\squeezetable
\caption{Experimental values of axial-vector couplings in
decuplet-to-octet processes.
}
\label{t07}
\begin{tabular}{
l
r@{.}l@{\,$\pm$\,}r@{.}l r@{.}l@{\,$\pm$\,}r@{.}l
r@{.}l@{\,$\pm$\,}r@{.}l r@{.}l@{\,$\pm$\,}r@{.}l
}
&
\multicolumn{4}{c}{$\Delta \rightarrow N \pi$} &
\multicolumn{4}{c}{$\Sigma^* \rightarrow \Lambda \pi$} &
\multicolumn{4}{c}{$\Sigma^* \rightarrow \Sigma \pi$} &
\multicolumn{4}{c}{$\Xi^* \rightarrow \Xi \pi$} \\
\tableline
$g$
 & $-2$ & 04 & 0 & 01  & $-$1 & 71 & 0 & 03
 & $-1$ & 60 & 0 & 13  & $-1$ & 42 & 0 & 04 \\
\end{tabular}
\end{table}


\begin{table}
\squeezetable
\caption{Best fitted parameters for the vector and axial-vector form
factors. $V_{ud}$ and $V_{us}$ in Fits A and B are inputs. Errors are from
the $\chi^2$ fit only, and do not include any theoretical uncertainties.}
\label{t09}
\begin{tabular}{
l
r@{.}l@{\,$\pm$\,}r@{.}l r@{.}l@{\,$\pm$\,}r@{.}l 
r@{.}l@{\,$\pm$\,}r@{.}l
r@{.}l@{\,$\pm$\,}r@{.}l
}
&
\multicolumn{4}{c}{Fit A} & \multicolumn{4}{c}{Fit B} &
\multicolumn{4}{c}{Fit C} & \multicolumn{4}{c}{Fit D}\\
\tableline
$V_{ud}$ &
0 & 9736 & 0 & 0010 & 0 & 9736 & 0 & 0010 & 0 & 9743 & 0 & 0009 & 0 & 9743 &
0 &  0009 \\
$V_{us}$ &
0 & 2196 & 0 & 0023 & 0 & 2196 & 0 & 0023 & 0 & 2194 & 0 & 0023 & 0 & 2194 & 0
& 0023 \\
$v_1$ &
\multicolumn{4}{c}{} & $-$0 & 03 & 0 & 04 & $-$0 & 03 & 0 & 04 & $-$0 & 02 &
0& 04 \\
$v_2$ &
\multicolumn{4}{c}{} & 0 & 05 & 0 & 03 & 0 & 05 & 0 & 03 & 0 & 05 & 0 & 03\\
$v_3$ &
\multicolumn{4}{c}{} & $-$0 & 01 & 0 & 01 & $-$0 & 01 & 0 & 01 & $-$0 & 01 & 0
& 01 \\
$a$ &
0 & 86 & 0 & 02 & 0 & 87 & 0 & 02 & 0 & 87 & 0 & 02 & 0 & 84 & 0 & 02\\
$b$ &
$-$0 & 16 & 0 & 03 & $-$0 & 18 & 0 & 03 & $-$0 & 18 & 0 & 03 & $-$0 & 12 & 0 &
03\\
$d$ &
$-$0 & 07 & 0 & 01 & $-$0 & 07 & 0 & 01 & $-$0 & 07 & 0 & 01 & $-$0& 03 & 0 &
01\\
$c_1$ &
$-$0 & 03 & 0 & 02 & $-$0 & 03 & 0 & 02 & $-$0 & 03 & 0 & 02 & $-$0 & 01 & 0 &
02\\
$c_2$ &
0 & 09 & 0 & 04 & 0 & 10 & 0 & 04 & 0 & 10 & 0 & 04 & 0 & 05 & 0 & 04 \\
$c_3$ &
$-$0 & 06 & 0 & 01 & $-$0 & 07 & 0 & 01 & $-$0 & 07 & 0 & 01 & $-$0 & 05 &
0 &01 \\
$c_4$ &
0 & 04 & 0 & 01 & 0 & 03 & 0 & 01 & 0 & 03 & 0 & 01 & 0 & 02 & 0 & 01\\
$F$ &
0 & 41 & 0 & 02 & 0 & 40 & 0 & 02 & 0 & 40 & 0 & 02 & 0 &  43 & 0 & 02\\
$D$ &
0 & 86 & 0 & 02 & 0 & 87 & 0 & 02 & 0 & 87 & 0 & 02 & 0 & 84 & 0 & 02\\
$3F-D$ &
0 & 37 & 0 & 08 & 0 & 34 & 0 & 08 & 0 & 34 & 0 & 08 & 0 & 46 & 0 & 08
\end{tabular}
\end{table}


\begin{table}
\squeezetable
\caption{Predicted form factors. Errors are from the $\chi^2$ fit only,
and do not include any theoretical uncertainties. $f_2$ has the same
values for fits A--D, and $f_1$ has the same values for fits B--D.}
\label{t10}
\begin{tabular}{
l
r@{.}l@{\,\,\,\,}r@{.}l@{\,\,\,\,}r@{.}l@{\,$\pm$\,}r@{.}l
r@{.}l@{\,\,\,\,}r@{.}l@{\,$\pm$\,}r@{.}l
r@{.}l@{\,$\pm$\,}r@{.}l
r@{.}l@{\,$\pm$\,}r@{.}l
}
&
\multicolumn{8}{c}{Fit A} & \multicolumn{6}{c}{Fit B} &
\multicolumn{4}{c}{Fit C} & \multicolumn{4}{c}{Fit D} \\
\tableline
Transition &
\multicolumn{2}{c}{$f_1$} & \multicolumn{2}{c}{$f_2$} &
\multicolumn{4}{c}{$g_1$} &
\multicolumn{2}{c}{$f_1$} & \multicolumn{4}{c}{$g_1$} &
\multicolumn{4}{c}{$g_1$} & \multicolumn{4}{c}{$g_1$} \\
\tableline
$n \rightarrow p$ &
1 & 00 & 1 & 85 & 1 & 269 & 0 & 001 & 
1 & 00 & 1 & 269 & 0 & 001 &
1 & 268 & 0 & 002 & 1 & 268 & 0 & 002 \\
$\Sigma^{\pm} \rightarrow \Lambda$ &
0 & 00 & 1 & 17 & 0 & 60 & 0 & 01 &
0 & 00 & 0 & 60 & 0 & 01 &
0 & 60 & 0 & 01 & 0 & 60 & 0 & 01 \\
$\Lambda \rightarrow p$ &
$-$1 & 22 & $-$1 & 10 & $-$0 & 89 & 0 & 01 &
$-$1 & 25$\pm$0.02 & $-$0 & 89 & 0 & 01 &
$-$0 & 89 & 0 & 01 & $-$0 & 89 & 0 & 01 \\
$\Sigma^- \rightarrow n$ &
$-$1 & 00 & 1 & 02 & 0 & 34 & 0 & 01 &
$-$1 & 04$\pm$0.02 & 0 & 34 & 0 & 01 &
0 & 34 & 0 & 01 & 0 & 34 & 0 & 01 \\
$\Xi^- \rightarrow \Lambda$ &
1 & 22 & $-$0 & 07 & 0 & 27 & 0 & 03 &
1 & 35$\pm$0.05 & 0 & 25 & 0 & 03 &
0 & 25 & 0 & 03 & 0 & 25 & 0 & 03 \\
$\Xi^-\rightarrow \Sigma^0$ &
0 & 71 & 1 & 31 & 0 & 73 & 0 & 02 &
0 & 79$\pm$0.03 & 0 & 72 & 0 & 02 &
0 & 72 & 0 & 02 & 0 & 76 & 0 & 02 \\
$\Xi^0\rightarrow \Sigma^+$ &
1 & 00 & 1 & 85 & 1 & 03 & 0 & 02 &
1 & 12$\pm$0.05 & 1 & 02 & 0 & 02 &
1 & 02 & 0 & 03 & 1 & 07 & 0 & 03 
\end{tabular}
\end{table}


\begin{table}
\squeezetable
\caption{Symmetry breaking for $f_1$. The ratio $f_1/f_1^{\rm SU(3)}$ is 
displayed. Errors are from
the $\chi^2$ fit only, and do not include any theoretical uncertainties.}
\label{t11}
\begin{tabular}{lccccc}
Transition &
Fit B, C, D & Anderson and Luty~\cite{and} & Donoghue {\it et 
al.}~\cite{don} & Krause~\cite{kra} &  Schlumpf~\cite{sch} \\
\tableline
$\Lambda \rightarrow p$ &
1.02$\pm$0.02 & 1.024 & 0.987 & 0.943 & 0.976 \\
$\Sigma^- \rightarrow n$ &
1.04$\pm$0.02 & 1.100 & 0.987 & 0.987 & 0.975 \\
$\Xi^- \rightarrow \Lambda$ &
1.10$\pm$0.04 & 1.059 & 0.987 & 0.957 & 0.976 \\
$\Xi^-\rightarrow \Sigma^0$ &
1.12$\pm$0.05 & 1.011 & 0.987 & 0.943 & 0.976 \\
$\Xi^0\rightarrow \Sigma^+$ &
1.12$\pm$0.05 & & & 
\end{tabular}
\end{table}


\begin{table}
\squeezetable
\caption{Theoretical predictions for decuplet-to-octet axial-vector 
couplings $g$. Errors are from
the $\chi^2$ fit only, and do not include any theoretical uncertainties.}
\label{t12}
\begin{tabular}{
l
r@{.}l@{\,$\pm$\,}r@{.}l r@{.}l
r@{.}l@{\,$\pm$\,}r@{.}l r@{.}l
r@{.}l@{\,$\pm$\,}r@{.}l r@{.}l
}
Transition &
\multicolumn{4}{c}{Fit A} & \multicolumn{2}{c}{$\chi^2$} &
\multicolumn{4}{c}{Fit B} & \multicolumn{2}{c}{$\chi^2$} &
\multicolumn{4}{c}{Fit C} & \multicolumn{2}{c}{$\chi^2$} \\
\tableline
$\Delta \rightarrow N$ & $-$2 & 03 & 0 & 01 & 0 & 3 & $-$2 & 04 & 0 & 01
& 0 & 2 & $-$2 & 04 & 0 & 01 & 0 & 2 \\
$\Sigma^* \rightarrow \Lambda$ & $-$1 & 78 & 0 & 02 & 5 & 7 & $-$1 & 77 &
0 & 02 & 3 & 7 & $-$1 & 77 & 0 & 02 & 3 & 7 \\
$\Sigma^* \rightarrow \Sigma$ & $-$1 & 49 & 0 & 07 & 0 & 7 & $-$1 & 55 &
0 & 07 & 0 & 1 & $-$1 & 55 & 0 & 07 & 0 & 1 \\
$\Xi^* \rightarrow \Xi$ & $-1$ & 38 & 0 & 04 & 0 & 8 & $-1$ & 39 & 0 & 04 
& 0 & 5 & $-$1 & 39 & 0 & 04 & 0 & 5 \\
\end{tabular}
\end{table}


\begin{table}
\squeezetable
\caption{Theoretical predictions for three $\Delta S = 0$ hyperon 
semileptonic decays and their contributions to the total $\chi^2$. The 
units of $R$ are 10$^{-3}$ s$^{-1}$ for neutron decay and 10$^6$ s$^{-1}$
for the remaining decays. The predictions for Fits C and D are the same as for
Fit B.}
\label{t13}
\begin{tabular}{
l
r@{.}l@{\,\,\,}r@{.}l r@{.}l@{\,\,\,}r@{.}l r@{.}l@{\,\,\,}r@{.}l
r@{.}l@{\,\,\,}r@{.}l r@{.}l@{\,\,\,}r@{.}l r@{.}l@{\,\,\,}r@{.}l
r@{.}l@{\,\,\,}r@{.}l r@{.}l@{\,\,\,}r@{.}l r@{.}l@{\,\,\,}r@{.}l
r@{.}l@{\,\,\,}r@{.}l r@{.}l@{\,\,\,}r@{.}l r@{.}l@{\,\,\,}r@{.}l
}
&
\multicolumn{8}{c}{$n \rightarrow p e^- \overline \nu_e$} &
\multicolumn{8}{c}{$\Sigma^+ \rightarrow \Lambda e^+ \nu_e$} &
\multicolumn{8}{c}{$\Sigma^- \rightarrow \Lambda e^- \overline \nu_e$} \\
\tableline
 &
\multicolumn{2}{c}{Fit A} & \multicolumn{2}{c}{$\chi^2$} &
\multicolumn{2}{c}{Fit B} & \multicolumn{2}{c}{$\chi^2$} &
\multicolumn{2}{c}{Fit A} & \multicolumn{2}{c}{$\chi^2$} &
\multicolumn{2}{c}{Fit B} & \multicolumn{2}{c}{$\chi^2$} &
\multicolumn{2}{c}{Fit A} & \multicolumn{2}{c}{$\chi^2$} &
\multicolumn{2}{c}{Fit B} & \multicolumn{2}{c}{$\chi^2$} \\
\tableline
$R$ &
1 & 13 & 0 & 9 & 1 & 13 & 0 & 9 &
0 & 23 & 0 & 1 & 0 & 23 & 0 & 1 &
0 & 39 & 0 & 0 & 0 & 39 & 0 & 0 \\
$\alpha_{e\nu}$ &
$-$0 & 08 & 0 & 3 & $-$0 & 08 & 0 & 3 &
$-$0 & 41 & 0 & 1 & $-$0 & 41 & 0 & 1 &
$-$0 & 41 & 0 & 1 & $-$0 & 41 & 0 & 1 \\
$\alpha_e$ &
$-$0 & 09 & 2 & 7 & $-$0 & 09 & 2 & 7 & 
\multicolumn{4}{c}{} & \multicolumn{4}{c}{} &
\multicolumn{4}{c}{} & \multicolumn{4}{c}{} \\
$\alpha_\nu$ &
0 & 99 & 0 & 1 & 0 & 99 & 0 & 1 &
\multicolumn{4}{c}{} & \multicolumn{4}{c}{} &
\multicolumn{4}{c}{} & \multicolumn{4}{c}{} \\
$A$ &
\multicolumn{4}{c}{} & \multicolumn{4}{c}{} &
\multicolumn{4}{c}{} & \multicolumn{4}{c}{} &
0 & 05 & 0 & 1 & 0 & 05 & 0 & 1 \\
$B$ &
\multicolumn{4}{c}{} & \multicolumn{4}{c}{} &
\multicolumn{4}{c}{} & \multicolumn{4}{c}{} &
0 & 88 & 0 & 2 & 0 & 88 & 0 & 2 \\
$g_1/f_1$ &
1 & 27 & \multicolumn{2}{c}{} & 1 & 27 & \multicolumn{2}{c}{} &
\multicolumn{4}{c}{} & \multicolumn{4}{c}{} &
\multicolumn{4}{c}{} & \multicolumn{4}{c}{}
\end{tabular}
\end{table}


\begin{table}
\squeezetable
\caption{Theoretical predictions for five $|\Delta S| = 1$ hyperon 
semileptonic decays and their contributions to the total $\chi^2$. The 
units of $R$ are 10$^{6}$ s$^{-1}$. The values for Fits C and D are the same as
for Fit B for the first three decay modes. For $\Xi^- \to \Sigma^0$ decay, Fit
C gives the same values as Fit B, and Fit D
gives $R=0.4$ and $g_1/f_1=0.96$.}
\label{t14}
\begin{tabular}{
l
r@{.}l@{\,\,\,\,}r@{.}l r@{.}l@{\,\,\,\,}r@{.}l r@{.}l@{\,\,\,\,}r@{.}l 
r@{.}l@{\,\,\,\,}r@{.}l r@{.}l@{\,\,\,\,}r@{.}l r@{.}l@{\,\,\,\,}r@{.}l
r@{.}l@{\,\,\,\,}r@{.}l r@{.}l@{\,\,\,\,}r@{.}l r@{.}l@{\,\,\,\,}r@{.}l
r@{.}l@{\,\,\,\,}r@{.}l r@{.}l@{\,\,\,\,}r@{.}l r@{.}l@{\,\,\,\,}r@{.}l
r@{.}l@{\,\,\,\,}r@{.}l r@{.}l@{\,\,\,\,}r@{.}l r@{.}l@{\,\,\,\,}r@{.}l
r@{.}l@{\,\,\,\,}r@{.}l r@{.}l@{\,\,\,\,}r@{.}l r@{.}l@{\,\,\,\,}r@{.}l
}
&
\multicolumn{8}{c}{$\Lambda \rightarrow p e^- \overline \nu_e$} &
\multicolumn{8}{c}{$\Sigma^- \rightarrow n e^-\overline \nu_e$} &
\multicolumn{8}{c}{$\Xi^- \rightarrow \Lambda e^- \overline \nu_e$} &
\multicolumn{8}{c}{$\Xi^-\rightarrow \Sigma^0 e^- \overline \nu_e$} &
\multicolumn{6}{c}{$\Xi^0\rightarrow \Sigma^+ e^- \overline \nu_e$} \\
\tableline
 &
\multicolumn{2}{c}{Fit A} & \multicolumn{2}{c}{$\chi^2$} &
\multicolumn{2}{c}{Fit B} & \multicolumn{2}{c}{$\chi^2$} &
\multicolumn{2}{c}{Fit A} & \multicolumn{2}{c}{$\chi^2$} &
\multicolumn{2}{c}{Fit B} & \multicolumn{2}{c}{$\chi^2$} &
\multicolumn{2}{c}{Fit A} & \multicolumn{2}{c}{$\chi^2$} &
\multicolumn{2}{c}{Fit B} & \multicolumn{2}{c}{$\chi^2$} &
\multicolumn{2}{c}{Fit A} & \multicolumn{2}{c}{$\chi^2$} &
\multicolumn{2}{c}{Fit B} & \multicolumn{2}{c}{$\chi^2$} &
\multicolumn{2}{c}{Fit A} & \multicolumn{2}{c}{Fit B} &
\multicolumn{2}{c}{Fit D} \\
\tableline
$R$ &
3 & 14 & 0 & 1 & 3 & 19 & 0 & 2 &
6 & 43 & 3 & 6 & 6 & 83 & 0 & 0 &
2 & 87 & 8 & 7 & 3 & 32 & 0 & 4 &
0 & 36 & 2 & 7 & 0 & 38 & 2 & 2 &
0 & 65 & 0 & 68 & 0 & 73 \\
$\alpha_{e\nu}$ &
$-$0 & 03 & 1 & 3 & $-$0 & 02 & 0 & 0 &
0 & 34 & 0 & 2 & 0 & 36 & 0 & 3 &
0 & 60 & 0 & 5 & 0 & 65 & 1 & 5 &
\multicolumn{4}{c}{} & \multicolumn{4}{c}{} &
$-$0 & 14 & $-$0 & 07 & $-$0 & 10 \\
$\alpha_e$ &
0 & 01 & 2 & 9 & 0 & 02 & 2 & 4 &
$-$0 & 63 & 1 & 2 & $-$0 & 61 & 0 & 8 &
\multicolumn{4}{c}{} & \multicolumn{4}{c}{} &
\multicolumn{4}{c}{} & \multicolumn{4}{c}{} &
$-$0 & 11 & $-$0 & 05 & $-$0 & 07 \\
$\alpha_\nu$ &
0 & 98 & 6 & 9 & 0 & 97 & 6 & 6 &
$-$0 & 35 & 4 & 0 & $-$0 & 35 & 3 & 8 &
\multicolumn{4}{c}{} & \multicolumn{4}{c}{} &
\multicolumn{4}{c}{} & \multicolumn{4}{c}{} &
1 & 00 & 1 & 00 & 1 & 00 \\
$\alpha_B$ &
$-$0 & 59 & 1 & 5 & $-$0 & 59 & 1 & 7 &
0 & 67 & 2 & 4 & 0 & 65 & 1 & 9 &
\multicolumn{4}{c}{} & \multicolumn{4}{c}{} &
\multicolumn{4}{c}{} & \multicolumn{4}{c}{} &
$-$0 & 52 & $-$0 & 56 & $-$0 & 55 \\
$A$ &
\multicolumn{4}{c}{} & \multicolumn{4}{c}{} &
\multicolumn{4}{c}{} & \multicolumn{4}{c}{} &
0 & 53 & 0 & 8 & 0 & 46 & 2 & 6 &
\multicolumn{4}{c}{} & \multicolumn{4}{c}{} &
0 & 71 & 0 & 75 & 0 & 73 \\
$B$ &
\multicolumn{4}{c}{} & \multicolumn{4}{c}{} &
\multicolumn{4}{c}{} & \multicolumn{4}{c}{} &
\multicolumn{4}{c}{} & \multicolumn{4}{c}{} &
\multicolumn{4}{c}{} & \multicolumn{4}{c}{} &
0 & 61 & 0 & 56 & 0 & 59 \\
$g_1/f_1$ &
0 & 73 & \multicolumn{2}{c}{} &
0 & 71 & \multicolumn{2}{c}{} &
$-$0 & 34 & \multicolumn{2}{c}{} &
$-$0 & 33 & \multicolumn{2}{c}{} &
0 & 22 & \multicolumn{2}{c}{} &
0 & 18 & \multicolumn{2}{c}{} &
1 & 03 & \multicolumn{2}{c}{} &
0 & 91 & \multicolumn{2}{c}{} &
1 & 03 & 0 & 91 & 0 & 96
\end{tabular}
\end{table}

\end{document}